\documentclass[a4paper,12pt]{article}

\usepackage{amsmath}\usepackage{amssymb}
\usepackage{latexsym}\usepackage{cite}

\usepackage[dvips]{graphicx}
\usepackage{pstricks}
\usepackage{bm}

\usepackage{graphicx}

\newcommand{\be}{\begin{equation}}
\newcommand{\ee}{\end{equation}}

\newcommand{\ka}{\kappa}

\def\beq{\begin{equation}}
\def\eeq{\end{equation}}

\def\al{\alpha}
\def\bt{\beta}

\def\ga{\gamma}
\def\de{\delta}

\def\ka{\kappa}

\def\te{\theta}

\def\lam{\lambda}
\def\Om{\Omega}

\def\sq{\sqrt}

\def\l{\left (}
\def\r{\right )}

\def\fr{\frac}
\def\la{\label}
\def\hs{\hspace}
\def\vs{\vspace}

\def\ran{\rangle}
\def\lan{\langle}
\def\ov{\overline}
\def\tl{\tilde}
\def\tm{\times}

\begin{document}

\begin{flushright}
OSU-HEP-08-03\\
NSF-KITP-08-51\\
April 12, 2008 \\
\end{flushright}

\vs{1.5cm}

\begin{center}
{\Large\bf

A New Extensions of MSSM: FMSSM}
\end{center}

\vspace{0.5cm}
\begin{center}
{\large
{}~S.~Nandi\footnote{E-mail: s.nandi@okstate.edu}{}~ and
{}~Zurab Tavartkiladze\footnote{E-mail: zurab.tavartkiladze@okstate.edu}
}
\vspace{0.5cm}

{\em Department of Physics and Oklahoma Center for High Energy
Physics, Oklahoma State University, Stillwater, OK 74078, USA }

\end{center}
\vspace{0.6cm}

\begin{abstract}


We propose an extension of the MSSM by adding vector like `matter'
fields with masses near the TeV scale. This extension allows us to
generate the masses of the bottom quark and tau lepton via radiative
corrections such that only up type Higgs doublet couples with quarks
and leptons. In addition to providing a natural explanation of the
hierarchies between $m_{b, \tau }$ and $m_t$, this new extension,
which we call FMSSM, allows the heavy sector of the MSSM Higgs
bosons to be essentially fermiophobic as well as gaugephobic.
Moreover, in this scenario there is no upper bound for the parameter
$\tan \beta $. FMSSM can be distinguished from the MSSM, and has
peculiar and unorthodox signals at the LHC, especially for the Higgs
sector.

\end{abstract}







\newpage

\section{Introduction}

All experimental results to date are in excellent agreement with the
predictions of the Standard Model (SM). One essential ingredient of
the SM, the existence of the Higgs boson as well as its interactions
with the SM particles, are yet to be experimentally established. It
is widely believed that the Higgs boson will be discovered at the
LHC (even possibly at the Tevatron). However, the most serious
theoretical drawback of the SM is the lack of explanation  why the
Higgs boson's mass is near electro-weak (EW) scale, and not at the
Planck scale, the so called hierarchy problem. The supersymmetric
(SUSY) extension of the SM is most compelling in the sense that it
solves the hierarchy problem naturally. Also the minimal SUSY
extension of the SM (MSSM) leads to the successful gauge coupling
unification. This gives a compelling ground for grand unification
\cite{Pati:1974yy}. Moreover, the MSSM  provides a viable candidate
for the dark matter. In addition, it has a rich spectrum of
particles, namely additional Higgs bosons, and the superpartners of
the gauge bosons and squarks and sleptons which can be explored at
the LHC.

In MSSM the Yukawa couplings are free parameters and should be
chosen in such a way as to obtain observed hierarchies between
fermion masses and CKM mixing angles. While the top quark mass is
close to the EW symmetry breaking scale, the masses of other members
of a third family - the bottom quark and tau lepton are smaller by
factor of $\sim 60$ and $\sim 140$ respectively. The MSSM parameter
$\tan \bt $ can give this mismatch, however, this parameter gets
constrained from various observables. This can be avoided, and also
$m_{b,\tau }/m_t\ll 1$ hierarchies can be explained if $m_b$ and
$m_{\tau }$ are generated by radiative corrections. Note that, the
Yukawa sector of MSSM also does not allow to have the fermiophobic
Higgses unlike, for instance, for the general two Higgs doublet
model \cite{Akeroyd:1995hg}. However, an extension providing
radiative bottom-tau mass generation may open such an interesting
possibility. Our new extension to the MSSM is motivated by this
philosophy.

To achieve all this, we extend the matter sector of MSSM by adding
vector-like quarks $D^c+\ov{D}^c$ and leptons $L+\bar L$ with masses
at the TeV scale. $D^c$ has the same quantum number as the usual
$d^c$, while $L$ has the same quantum number as the usual leptonic
doublet $l$. The top quark has  tree level renormalizable coupling
with the up type Higgs, however, the bottom quark and the tau lepton has no
such tree level coupling with the down type Higgs. Instead they have
only tree level  coupling with the down type Higgs via $D^c$ and $L$
respectively, allowing their mass generation at the one loop level
after SUSY breaking.
Besides this, this model
allows the possibility of having  essentially a fermiophobic heavy
Higgs sector which is not possible in the MSSM. The phenomenology of
this model is very different from the MSSM. In particular, the heavy
Higgs signals of fermiophobic MSSM (FMSSM) at the LHC is
significantly different from those of MSSM.

\section{Formalism of the model: FMSSM}\label{sect:FMSSM}

We first consider the third generation of quarks and leptons, which
in MSSM have the strongest couplings to the Higgs doublets. The
quark superfields are $q_3, u^c_3, d^c_3$ while $l_3$ and $e^c_3$
are the lepton superfields. As in the MSSM, our model has two Higgs
doublet
superfields, $\hat{h}_u$ and $\hat{h}_d$. In the MSSM, $\hat{h}_u$
couples to the up type quark, while $\hat{h}_d$ couples to the down
type quarks and charged leptons. In our model, similar to the MSSM, the top quark mass
is generated through the renormalizable Yukawa superpotential
term
\beq W_{up}=\lam_tq_3u^c_3\hat{h}_u~.
\la{Yt}
\eeq
In difference from MSSM, in our model, $\hat{h}_d$ does not directly
couple to the light matter. The masses of the bottom quark and the
tau lepton will be generated via radiative corrections. Namely,
after SUSY breaking, at the one loop level, the operators
$q_3d^c_3h_u^{\dag }$ and $l_3e^c_3h_u^{\dag }$ will be induced,
where $h_u$ denote the scalar component of the superfield
$\hat{h}_u$ and remaining states here stand for denoting fermionic
components. Since these coupling will be suppressed by loop factors,
the natural explanation of the hierarchies $\fr{m_b}{m_t},
\fr{m_{\tau }}{m_t}\ll 1$ is provided. Moreover, what is perhaps
most interesting, only $h_u$ doublet couples to the fermions. This
will allow us to have an essentially fermiophobic  as well as
gaugephobic heavy Higgs system ($H, A$ and $H^{\pm }$) as we will see shortly.

To generate the couplings mentioned above, we extend the fermionic
matter sector of the MSSM by adding the vector like matter
superfields $D^c+\ov{D}^c$ and $L+\bar L$. Transformation properties
of $D^c$ and $L$ under $SU(3)_c\tm SU(2)_L\tm U(1)_Y$ coincide with
transformations of $d^c$ and $l$ respectively. Thus, these
introduced states effectively constitute $SU(5)$ complete multiplets
$\bar 5+5$ [with $(D^c, L)\subset \bar 5$ and $(\ov{D}^c, \bar
L)\subset 5$]. As we have mentioned, $\hat{h}_d$ does not couple
directly with light matter, and the states $D^c,L$ should do the job
for the generation of $m_{b,\tau }$. For this, specific
superpotential as well as soft SUSY breaking terms should be
introduced. At the same time, absence of any interaction  should be
justified. For this purpose we will use $R$-symmetry. The $\mu $ and
$B_{\mu }$ terms will be generated after SUSY breaking.
Since the breaking of $R$-symmetry is an essential ingredient of the SUSY
breaking \cite{Nelson:1993nf}, it is advantageous to use  $R$-symmetry also
for other phenomenological purposes. In our construction we will follow this strategy.

For the usual MSSM fields, and for the new fields that we introduce,
 we make the following $R$-charge assignment.
$$
R(q)=R(u^c)=R(e^c)=R(D^c)=R(\ov{D}^c)=R(L)=R(\bar L)=1~,
$$
\beq
R(d^c)=R(l)=r~,~~~~~R(\hat{h}_u)=R(\hat{h}_d)=0~,
\la{r-charges}
\eeq
where $r$ is some phase (undetermined for time being). Here we consider family independent
 $R$-symmetry. We demand that the Lagrangian to be invariant under
 this R symmetry.
This assignment is compatible with the superpotential term of Eq. (\ref{Yt}). In addition, the
following superpotential couplings are allowed
\beq
W'=\lam_Dq_3D^c\hat{h}_d+\lam_LLe^c\hat{h}_d+M_DD^c\ov{D}^c+M_LL\bar L~.
\la{extW}
\eeq
As we see, the direct coupling of $h_d$ with the light matter is forbidden. Also, the $\mu $ term is not
allowed in the superpotential. $\mu $ and $B_{\mu }$ terms, as we have mentioned, will be generated after
SUSY breaking by the higher order operators, similar to the proposal of Ref. \cite{Giudice:1988yz}. Thus we introduce two spurion superfields $X$ and $Y$ with
\beq
\lan X\ran =M^{(X)}+\te^2m M_{\rm Pl}~,~~~~~\lan Y\ran =M^{(Y)}+\te^2m M_{\rm Pl}~,
\la{spur}
\eeq
and $R$-charges
\beq
R(X)=0~,~~~~R(Y)=r-1~.
\la{RXY}
\eeq
As we see, the SUSY breaking is by $F$-terms and the $R$-symmetry is also broken by the VEVs of (\ref{spur}).
We assume that a hidden sector is arranged is such a way that the configuration (\ref{spur}) is insured. Note also
that there should be included the constant part of the superpotential \cite{Chamseddine:1982jx} in order to be able to set the cosmological constant
to zero and get the Minkowski vacuum. This constant superpotential, breaking the $R$-symmetry explicitly, avoids pseudogoldstones
($R$-axions), and renders all the phenomenology discussed below intact.
In (\ref{spur}) $M_{\rm Pl}\simeq 2.4\cdot 10^{18}$~GeV is the reduced Planck mass, while $m\sim 1$~TeV\footnote{Here, for simplicity, we assume gravity
mediated SUSY breaking scenario. However, in this framework different SUSY breaking scenarios can work as well.}.
 Through the couplings
$$
\int d^4\te \fr{X^{\dag }}{M_{\rm Pl}}\hat{h}_u\hat{h}_d~,~~~\int d^4\te \fr{X^{\dag }X}{M_{\rm Pl}^2}\hat{h}_u\hat{h}_d~,
$$
after substituting the VEVs of Eq. (\ref{spur}), we have $\mu \sim
m$ and $B_{\mu }\sim m^2$. Furthermore, the operators $\int d^2\te
\fr{X}{M_{\rm Pl}}W_aW_a$ (where $W_a$ is the chiral gauge
superfield) generate the gaugino masses $M_a\sim m$. The trilinear
soft SUSY breaking terms will be generated from the operators $\int
d^2\te \fr{X}{M_{\rm Pl}}W$ where under $W$ we denote all the terms
which are included in the superpotential. The mass$^2$ soft SUSY
breaking terms are induced through the following operators
$$
\int d^4\te \fr{X^{\dag }X}{M_{\rm Pl}^2}f^{\dag }f~,~~~{\rm with}~~f=(q, u^c, d^c, l, e^c, D^c, \ov{D}^c, L, \bar L, \hat{h_u}, \hat{h}_d)~,
$$
\beq
\int d^4\te \l \fr{X^{\dag }Y}{M_{\rm Pl}^2}(D^cd^{c\dag }_3+Ll_3^{\dag })+\fr{XY^{\dag }}{M_{\rm Pl}^2}(D^{c\dag }d^c_3+L^{\dag }l_3)\r~.
\la{m2-ops}
\eeq
With these we will  have the following terms
$$
m^2_{\tl{q}_3}|\tl{q}_3|^2+m^2_{\tl{d}^c_3}|\tl{d}^c_3|^2+m^2_{\tl{l}_3}|\tl{l}_3|^2+m^2_{\tl{e}^c_3}|\tl{e}^c_3|^2+
$$
\beq (m^2_{dD}\tl{d}^{c*}_3\tl{D}^c+m^2_{lL}\tl{l}^{*}_3\tl{L}+{\rm
h.c.}) \la{sbt} \eeq with $m^2_{\tl{q}_3}\sim m^2_{\tl{d}^c_3}\sim
m^2_{\tl{l}_3}\sim m^2_{\tl{e}^c_3}\sim m^2_{dD}\sim m^2_{lL}\sim
m^2$. (We list here only  soft mass terms relevant for $m_{b,\tau }$
generation and omit terms like $m^2_{\tl{D}^c}|\tl{D}^c|^2$,
$m^2_{\tl{L}}|\tl{L}|^2$, etc.).

\begin{figure}[t]
\begin{center}
\hs{-1cm}
\resizebox{1.17\textwidth}{!}{
 \hs{-2.5cm} \vs{-13cm}\includegraphics{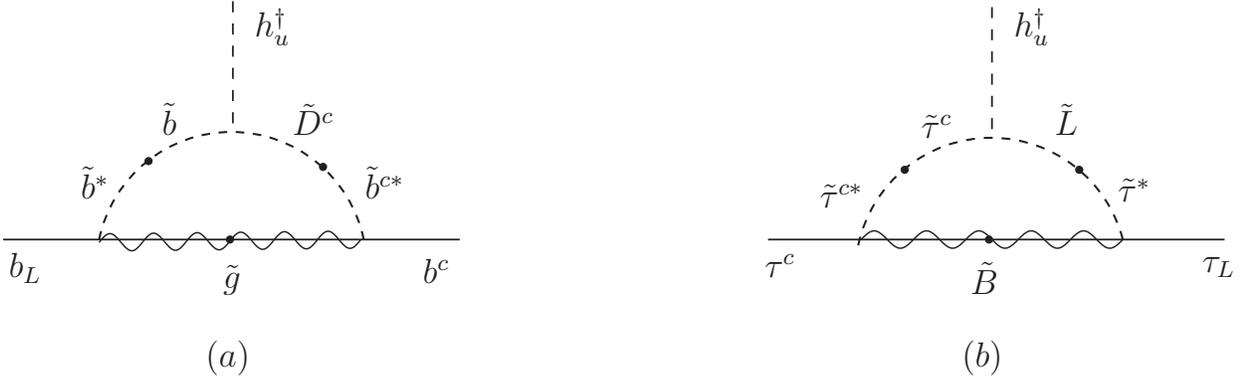}
}
\vs{-28cm}
\caption{Diagrams generating bottom quark and tau lepton masses.}
\label{fig:1}       
\end{center}
\end{figure}

It is easy to see that the superpotential term in Eqs. (\ref{Yt}),
(\ref{extW}) together with the soft breaking terms in Eq.
(\ref{sbt}), at 1-loop level,  generate the operators
\beq \lam_bq_3d^c_3h_u^{\dag
}~,~~~~~\lam_{\tau }l_3e^c_3h_u^{\dag }~, \la{Ybtau} \eeq with
$$
\lam_b =\fr{\lam_D\al_3}{4\pi }\fr{8}{3}\mu \ka M_{\tl{g}}\fr{m^2_{dD}}{m^4_{\tl{q}_3}}I_q~,~~~~~~~~~~
\lam_{\tau }=\fr{\lam_L\al_1}{4\pi }\fr{3}{5}\mu \ka M_{\tl{B}}\fr{m^2_{lL}}{m^4_{\tl{e}^c_3}}I_l
$$
$$
I_q= I\l \fr{M_{\tl{g}}^2}{m^2_{\tl{q}_3}}, \fr{M_D^2+m^2_{\tl{D}^c}}{m^2_{\tl{q}_3}},\fr{m^2_{\tl{d}^c_3}}{m^2_{\tl{q}_3}},\fr{m^4_{dD}}{m^4_{\tl{q}_3}}\r ~~~~~~
I_l= I\l \fr{M_{\tl{B}}^2}{m^2_{\tl{e}^c_3}},\fr{M_L^2+m^2_{\tl{L}}}{m^2_{\tl{e}^c_3}},\fr{m^2_{\tl{l}_3}}{m^2_{\tl{e}^c_3}},\fr{m^4_{lL}}{m^4_{\tl{e}^c_3}}\r,
$$
\beq
I(a, b, c, d)=\int_0^{\infty }\hs{-0.2cm}\fr{tdt}{t+1} \hs{0.1cm}\fr{1}{t+a} \hs{0.1cm}
 \fr{1}{(t+b)(t+c)-d}~.
\la{lam-btau}
\eeq
The corresponding diagrams are shown in Fig. \ref{fig:1}. The vortex involving $h_u^{\dag }$ emerges from the potential
term $|F_{h_d}|^2=|\lam_D\tl{q}_3\tl{D}^c+\lam_L\tl{L}\tl{e}^c_3+\ka \mu h_u|^2
\to \mu \ka (\lam_D\tl{q}_3\tl{D}^ch_u^{\dag }+\lam_L\tl{L}\tl{e}^c_3h_u^{\dag })$, where $\ka $ is some dimensionless
constant of the order of one. The operator, $\lam^{(1)}q_3d^c_3h_d$ (with $\lam^{(1)}\sim \lam_b\fr{\lan Y\ran }{M_{\rm Pl}}$) generated by the loops, will be  suppressed for small values of $\fr{\lan Y\ran }{M_{\rm Pl}}=\fr{M^{(Y)}}{M_{\rm Pl}}$. To be more specific,
this coupling induces the correction to the bottom quark mass $\de m_b\sim m_b\fr{\lan Y\ran }{M_{\rm Pl}}\fr{1}{\tan \bt }$.
Already for $\tan \bt \sim 10$, this correction will be negligible with $\fr{\lan Y\ran }{M_{\rm Pl}}\stackrel{<}{_\sim}0.1$.
Moreover, the $\lam^{(1)}$ coupling would induce corrections to the matter-Higgs interactions. For instance, the $b\bar b$ interaction with heavy CP even neutral Higgs ($H$) will receive the correction $\de \lam_{b\bar bH}\sim \fr{m_b}{\sq{2}v}\fr{\lan Y\ran }{M_{\rm Pl}}$, which still will
not give anything new if $\fr{\lan Y\ran }{M_{\rm Pl}}\stackrel{<}{_\sim}\fr{m_h^2}{m_H^2}\stackrel{<}{_\sim}\fr{1}{50}$.
The latter ratio controlls the suppression factor of the heavy Higgs fields' interaction with the matter (see discussion below).
The value of $\lan Y\ran \stackrel{<}{_\sim}M_{\rm Pl}/50$ is not unnatural. Since the VEV $\lan Y\ran $ should depend on the specific superpotential
couplings, with the dimensionless couplings of the order of $1/3-3$ and mass terms with values $M_{\rm Pl}/5$ say, the suppression factor
$1/50$ can be easily emerged without any fine tunings. Thus, with such suppressed $\lan Y\ran ~(\stackrel{<}{_\sim}M_{\rm Pl}/50)$,
the operator $\lam^{(1)}q_3d^c_3h_d$ becomes phenomenologically unimportant.

%
%
%
Note that if the additional vector-like particle ($D^c$ and L)
masses are in the TeV scale, which we take to be the SUSY breaking
scale, we obtain the  values of the masses of the bottom quark and
the tau lepton in their observed range (with appropriate choices for
the couplings $\lam_D$, $\lam_L\sim 1$ and soft masses). The mixed
soft terms $m^2_{dD}, m^2_{lL}$ in (\ref{sbt}) are important for
generating these operators. Since these off diagonal conversions
involve only third generation and the heavy vector like states, they
will not contribute to the FCNC. Thus with our explanation of why
the bottom quark and tau are so much lighter compared to the top
quark, we expect a vector-like quarks and leptons at the TeV scale.
The vector-like quark, being a color triplet, can be copiously pair
produced at the LHC. Thus our proposed model can be easily tested.
We will discuss the phenomenological implications of our model in
section \ref{sect:phen}.

Now we discuss the coupling of the mass eigenstate Higgs bosons with
the 3rd family fermions in our model. We have three neutral physical
Higgs bosons $(h, H, A)\equiv \phi $ where $h$ is lightest CP even
Higgs, $H$ is heavy($\sim 1$~TeV) CP even Higgs while $A$ is CP odd. From this and
the Yukawa couplings given in (\ref{Yt}) and (\ref{Ybtau}), we can
determine their interaction with third family fermions:
$$
{\cal L}_{f_3f_3\phi }= \fr{m_t}{\sq{2}vs_{\bt }}\left [ \bar tt
(c_{\al }h+s_{\al }H) +\bar ti\ga_5tc_{\bt }A\right ]
+\fr{m_b}{\sq{2}vs_{\bt }}\left [\bar bb (c_{\al }h+s_{\al }H) +\bar
bi\ga_5bc_{\bt }A\right]
$$
\beq +\fr{m_{\tau }}{\sq{2}vs_{\bt }}\left [\bar{\tau }\tau (c_{\al
}h+s_{\al }H) +\bar{\tau }i\ga_5\tau c_{\bt }A\right ]~.
\la{f3-f3-phi-int} \eeq The angle $\alpha$ is the mixing angle
between the two neutral CP-even Higgs bosons, h and H.

The interaction of the physical charged Higgs $H^{\pm }$ with the
3rd family fermions is given by
 \beq {\cal L}_{f_3f_3H^{\pm
}}=\fr{m_t}{v\tan \bt }b_Lt^cH^{+}+\fr{m_b}{v\tan \bt }\l
t_Lb^c+\fr{m_{\tau }}{m_b}\nu_{\tau }\tau^c\r H^{-}+ {\rm h.c}~.
\la{f3-f3-phi-int} \eeq Here $v\simeq 174$~GeV and the angles $\al ,
\bt $ are related as follows
 \beq
 \fr{\sin 2\al }{\sin 2\bt }=-\fr{m_H^2+m_h^2}{m_H^2-m_h^2}~,~~~~
 \fr{\tan 2\al }{\tan 2\bt }=\fr{m_A^2+m_Z^2}{m_A^2-m_Z^2}~,~~~
 0<\bt <\fr{\pi }{2}~,~~-\fr{\pi }{2}<\al <0~.
 \la{al-bt}
 \eeq
For $m_A, m_H\gg m_Z, m_h$, the relation between the angles
$\al $ and $\bt $ becomes
 \beq
 \alpha =-\fr{\pi }{2}+\bt +{\cal O}\l \fr{m_h^2}{m_H^2}, \fr{m_Z^2}{m_A^2}\r ~.
 \eeq
 Note that, in this limit, the coupling of the light neutral Higgs,
 $h$, to the EW gauge bosons,which is proportional to $\sin(\beta - \alpha )$
 becomes close to the SM values, while the coupling of the heavy Higgs,
 H, which is
 proportional to $\cos(\beta - \alpha)$, is suppressed($\sim \fr{m_h^2}{m_H^2}, \fr{m_Z^2}{m_A^2}$).
In this limit, the Yukawa couplings of the lightest neutral CP
even Higgs ($h$) with the 3rd family of matter fermions become:
\beq
\lam_{t\bar th}\simeq \fr{m_t}{\sq{2}v}~,~~~~
\lam_{b\bar bh}\simeq \fr{m_b}{\sq{2}v}~,~~~~
 \lam_{\tau \bar{\tau }h}=\fr{m_{\tau }}{\sq{2}v}~.
 \eeq
 Thus the Yukawa couplings of the light Higgs becomes identical
  to those in the SM.
 The coupling of the bottom quarks and the
 tau lepton is different from MSSM. These are not proportional to
 $\sq{1+\tan^2\bt }$ as in the MSSM. This is because, in our model, the bottom
 quark and the tau lepton does not couple to $h_d$ directly. This is
 one distinctive feature of our model compared to the MSSM.

 The couplings of the heavy Higgs sector to the fermions are
 given by

$$\lam_{t\bar tH}\simeq \lam_{t\bar tA} =\fr{\lam_{b\bar
tH^{+}}}{\sq{2}}=\fr{m_t}{\sq{2}v}\fr{1}{\tan \bt }~,~~~~~~
\lam_{b\bar bH}\simeq \lam_{b\bar bA} =\fr{\lam_{t\bar
bH^{-}}}{\sq{2}}=\fr{m_b}{\sq{2}v}\fr{1}{\tan \bt }~,
$$
\beq \lam_{\tau \bar{\tau }H}\simeq \lam_{\tau \bar{\tau }A}
=\fr{\lam_{\nu_{\tau }\bar{\tau }H^{-}}}{\sq{2}}=\fr{m_{\tau
}}{\sq{2}v}\fr{1}{\tan \bt }~. \la{hff-coupl} \eeq

Note that
the superpotential couplings $q_3d^c_3\hat{h}_d$ and $l_3e^c_3\hat{h}_d$ are forbidden thanks to  the $R$-symmetry.
By the same token soft SUSY breaking terms
$\tl{q}_3\tl{D}^ch_d$ and $\tl{e}^c_3\tl{L}h_d$ are absent at tree level.
With $\tan \bt \to \infty $, which means that $\lan h_u^0\ran
=174$~GeV and $\lan h_d^0\ran \to 0$, all  couplings in (\ref{hff-coupl}) vanish and
$H$, $A$ and $H^{\pm }$ decouple from the third
generation. Thus, in this limit, they become fermiophobic
or quasi-fermiophobic.
The latter case can be realized if light
generations will couple to $h_d$. This will depend on how we
construct the Yukawa sector for the light families. However, in any
case those interactions with Higgses would be strongly suppressed by tiny
Yukawa couplings.
As far as the very large values of the $\tan \bt $ (or equivalently $\lan h_d^0\ran \ll \lan h_u^0\ran$) is concerned
this regime is quite realizable. The VEV $\lan h_d^0\ran $ depends on the values of soft mass squires $m^2_{h_d}$, $m^2_{h_u}$ and also
on the $B$-term. By proper selection of these parameters the regime $\tan \bt \gg 1$ can be realized.
As mentioned earlier, gauge interactions of the heavy Higgses
with $WW$ and $ZZ$ are also suppressed in this limit. Thus, their
productions will be highly suppressed via gluon-gluon fusion or the
intermediate vector boson fusion, which are the two dominant
mechanism for producing these Higgs bosons.

We will discuss the phenomenological
implications of these for the detection of the Higgs bosons at LHC
in section \ref{sect:phen}.

We now discuss the couplings of the Higgs bosons with the new
vector-like fermions, $D^c$ and $L$. Note that  $\hat{h}_d$ couples
with third generation in combination with heavy states $D^c$ and $L$
(see Eq. (\ref{extW})). From these couplings it is straightforward
to derive the Higgs interactions involving one heavy fermion. They
are given as follows
$$
{\cal L}_{f_3F\phi }=\fr{\lam_D}{\sq{2}}\l b_LD^c
+\fr{\lam_L}{\lam_D}\tau^cE^{(L)}\r\l -s_{\al }h+c_{\al}H-is_{\bt
}A\r +{\rm h.c.}~
$$
\beq {\cal L}_{f_3FH^{\pm }}=\lam_Ds_{\bt }\cdot
t_LD^cH^{-}+\lam_Ls_{\bt }\cdot \tau^cN^{(L)}H^{-}+{\rm h.c.}
\la{DL-h-int} \eeq where $E^{(L)}$ and $N^{(L)}$ denote the charged
and neutral components of $L$ (e.g. $L=(N^{(L)},E^{(L)})$). Note
that in the limit, $m_A, m_H, m_H^{+}\gg m_Z, m_h$, the coupling of
$h$ with $D^c$ and b vanishes, while the coupling of $H$ and $A$
become $\approx s_{\bt }\lam_{D,L} /\sq{2}$. Thus, if $D^c$ and $L$
states are produced at the LHC, and they are heavier than $H$, they
will decay to the heavy Higgs and the bottom quark. On the other
hand, if $D^c$ is lighter than $H$, then it will decay to the
lighter Higgs $h$ and $b$ with a coupling $\approx c_{\bt
}\lam_D/\sq{2}$ (which gets suppressed for large values of $\tan \bt
$).

Before closing this section, we give the matter-smatter interaction
with Higgsinos. From the superpotential coupling (\ref{Yt}), we have
the following up type higgsino ($\widetilde{h}_u$) Yukawa coupling
with the quark and squark: \beq \fr{m_t}{vs_{\bt }}\l
\tl{q}_3t^c+q_3\tl{t}^c\r \widetilde{h}_u+{\rm h.c.}
\la{q-sq-Uhiggsino1} \eeq Since there is no superpotential coupling
of $\hat{h}_d$ with light matter,
 the combinations like $\tl{q}_3b^c$, $\tl{l}_3\tau^c$ etc.
do not couple with $\widetilde{h}_d$. They receive couplings with
 $\widetilde{h}_u$ through 1-loop diagrams (similar to ones given
  in Fig. \ref{fig:1}):
$$
\lam_{\tl{q}_3b\widetilde{h}_u}\tl{q}_3^*\bar b^c\widetilde{h}_u+
\lam_{\tl{\tau }l_3\widetilde{h}_u}\tl{\tau}^{c*}\bar l_3\widetilde{h}_u+{\rm h.c.}
$$
\beq
{\rm with}~~~~\lam_{\tl{q}_3b\widetilde{h}_u}\simeq \fr{m_{\tau }}{vs_{\bt }}\fr{\lam_D}{3\lam_L}\fr{I'}{I_l}~,~~~~
\lam_{\tl{\tau }l_3\widetilde{h}_u}\simeq \fr{m_{\tau }}{vs_{\bt }}\fr{1}{2}\l \fr{I''}{I_l}+5\fr{M_{\tl{W}}}{M_{\tl{B}}}\fr{I'''}{I_l}\r ~,
\la{f-sf-Uhiggsino2}
\eeq
where $I',I'', I'''$ are loop integrals defined similar to those given
 in Eq. (\ref{lam-btau}).

At one loop level, the couplings of $\tl{q}_3t^c$,  $q_3\tl{t}^c$
with down type Higgsino $\widetilde{h}_d$ are induced:
$$
\lam_{\tl{q}_3t\widetilde{h}_d}\tl{q}_3^*\bar t^c\widetilde{h}_d+
\lam_{q_3\tl{t}\widetilde{h}_d}\bar q_3\tl{t}^c\widetilde{h}_d+{\rm h.c.}
$$
\beq
{\rm with}~~~~~\lam_{\tl{q}_3t\widetilde{h}_d}\sim \lam_{q_3\tl{t}\widetilde{h}_d}\sim \fr{1}{16\pi^2}\fr{m_t}{vs_{\bt}}~.
\la{q-sq-Dhiggsino}
\eeq
Note once more that $\widetilde{h}_d$ has no Yukawa interactions
with $\tl{q}_3b^c, q_3\tl{b}^c$, $\tl{l}_3\tau^c$ and $l_3\tl{\tau }^c$.
Moreover, existing higgsino couplings in (\ref{q-sq-Uhiggsino1})-(\ref{q-sq-Dhiggsino}) are rather insensitive to the values of
$\tan \bt(\stackrel{>}{_\sim }2) $. This fact will allow to have no
 upper bound on $\tan \bt $ and relaxes constrains on some SUSY parameters.
We will discuss the phenomenological implications
of this in more detail in the next section.

\section{Phenomenological Implications}\label{sect:phen}

We now briefly discuss some of the phenomenological implications
of the model. The model has following major phenomenological implications.

\vs{0.25cm}

 ~{\bf (i)~New vector-like quarks and leptons in the TeV scale}

\vs{0.25cm}

 Within our scenario the bottom quark and the tau lepton masses are
 generated at 1-loop level. This gives natural way of explaining their suppressed
 values in comparison of $m_t$. In order the mechanism to work, we need new
 vector like states near TeV scale. Therefore,   we expect to
 see a vector-like quarks $D^c+\ov{D}^c$, and a vector-like  leptons $L+\ov{L}$
 at the TeV scale. The  $D^c$ under $SU(3)_C\tm SU(2)_L\tm U(1)_Y$ transforms as $(\bar 3,1,-\fr{2}{\sq{60}})$,
 where $U(1)_Y$ hypercharge is taken with $SU(5)$ normalization.
  The transformation properties of $L$ state is $(1,2,\fr{3}{\sq{60}})$.
   Since $D^c$ has strong
 interactions, it can be pair produced at the LHC up to a mass of
 about 2 TeV. If $D^c$ is heavier than $H$, then it will decay
 dominantly to a b quark and $H$ (see Eq. (\ref{DL-h-int})). It can also decay to
 the t-quark and the charged Higgs, although this decay mode will be
 somewhat suppressed compared to $D^c \rightarrow b H$. So the
 dominant final state signal from the $D^c{D^c}^{\star }$ pair
 productions will depend on the decay mode of $H$ (${D^c}^{\star }$ should
  not be confused with $\ov{D}^c$, the latter is a `mirror' of $D^c$).
  The coupling
 for the $H \rightarrow t \bar{t}$ decay is $\l \frac{m_t}{\sqrt{2} v \tan \beta}+\de_{t\bar tH}\r $
(where $\de_{t\bar tH}\sim 10^{-2}$ is coming  by the 1-loop induced operator $q_3u^c_3h_d^{\dag }$
\cite{Hempfling:1993kv, Hall:1993gn}, in analogy of the diagram of Fig. \ref{fig:1}a),
  whereas the coupling for $H \rightarrow{WW}$ is
$\frac {g}{2}$ $ \frac{{m_h}^2}{{M_H^2}}$. Thus the final state
 signals from the $D^c{D^c}^{\star }$ pair productions will be
 $b \bar{b}$ $t \bar{t}$ or $b \bar{b} W W$, depending on the
 relative values of $1/\tan \beta$ and $\frac{{m_h}^2}{{M_H^2}}$.
 (Note that with $\tan \bt \to \infty $ the $Ht\bar t$ coupling saturates
 to $\de_{t\bar tH}\sim 10^{-2}$ and
 can be comparable with $HWW$ coupling).
 The cross section for the $D^c{D^c}^{\star }$ production at the
 LHC (with $\sq{s}=14$~TeV) is about $100$~fb \cite{Han:2008gy}. So this could be an observable signal even
 at the early runs of the LHC with luminosity of few ${\rm fb}^{-1}$.

\vs{0.25cm}

 ~{\bf (ii)~Different Higgs signals compared to what we expect in the MSSM}

\vs{0.25cm}

 In our model, in the limit of large $\tan \beta$ and $m_{H,A}\gg m_{Z,h}$, the gauge coupling
as well as the fermionic coupling of the light Higgs h is
essentially like the SM Higgs. One major difference with MSSM is
that the h coupling to $b \bar{b}$ and $\tau \bar{\tau}$ is not $\tan
\beta$ enhanced as in the MSSM, since $h_d$ does not couple directly
to the down type quarks and to the charged leptons in our model. Thus there is no
restriction on $\tan \beta$ in our model, either from the perturbativity or unitarity
of the $h b \bar{b}$ coupling. Neither  the enhancement of the rare
processes like $b\to s\ga $ \cite{Hall:1993gn, Carena:1998gk}
and $B_0 \rightarrow \mu^+ \mu^-$ \cite{Babu:1999hn} occurs by increase of $\tan \bt $.

In the SM or in MSSM, the pair production for the Higgs bosons have
very small cross section, and most likely, will not be observable at
the LHC. In our model, because a pair of the Higgs bosons can be
produced from the decays of the $D^c{D^c}^{\star }$ pairs, and this
$D^c{D^c}^{\star }$ are produced via strong interaction, the double
Higgs production has a sizable cross section at the LHC. For
example, for the $D^c $ mass of $1$~TeV, this cross section is about
$100$~fb, while for a $800$~GeV $D^c$ mass, the cross section is
$400$~fb. This double Higgs production will be the heavy Higgs pair
if the $D^c$ is heavier than $H$. However, if $H$ is heavier than
$D^c$, then this double Higgs production  \cite{Djouadi:1999hh} will
be the light Higgs pair, with anomalously large cross section
compared to that expected in the SM, or in the MSSM.

\vs{0.25cm}

 ~{\bf (iii)~SUSY signals}

\vs{0.25cm}

Our model will have different signals also with respect to the
detection or/and the production of SUSY particles. For example, as
was noted in previous section, the higgsino-fermion-sfermion
interactions are practically insensitive to the value of $\tan \bt
$. Therefore, any process which involve the neutralino or chargino
is not enhanced even for very large values of $\tan \bt $.

For example, after gluino pair production at LHC, each gluino decay
in lightest squark ($\tl{q}_l$) and corresponding quark: $\tl{g}\to
\tl{q}_l^*q$ provided that $M_{\tl{g}}>m_{\tl{q}_l}$. With
$\tl{q}_l=\tl{b}^c$ and the lightest SUSY particle (LSP) being the
neutralino ($\widetilde{\chi }^0_1$), the
 subsequent decay of $\tl{b}^c$ is through the channel
$\tl{b}^c\to b^c\widetilde{\chi }^0_1$. This decay is only due to gauge
 coupling $g_1$ and completely independent
of Yukawa interactions [see (\ref{q-sq-Uhiggsino1})-(\ref{q-sq-Dhiggsino})].
Thus, in this case, the process
$\tl{g}\tl{g}\to b\bar b+\slash \hs{-0.26cm}E$ is insensitive to the
value of $\tan \bt $.

Moreover, for example,  if the LSP is $\widetilde{\chi }^0_1$, then the
 rare decay of the next heavier neutralino
$\widetilde{\chi }^0_2\to \widetilde{\chi }^0_1\ga $ is not sensitive
 to $\tan \bt $ unlike the
MSSM, in which some restrictions on SUSY parameters should be
imposed \cite{Baer:2005jq}. Thus many of the constraints in MSSM do
not apply in our model.

\vs{0.25cm}

 ~{\bf (iv)~Relaxing restrictions from neutralino cold dark matter}

\vs{0.25cm}

Assuming that LSP is the $\widetilde{\chi }^0_1$ neutralino, the relic
 density $\Om_{\chi }h^2$ depends on
neutralino couplings. For instance, in MSSM, the $\tl{\tau }$ being
the  NLSP, the co-annihilation processes are sensitive to the value
of $\tan \bt $. These processes, in different
 scenarios, constrain the value of $\tan \bt $
and different SUSY parameters \cite{Ellis:2001msa},
\cite{Baer:2005jq}. In our scenario, as we have seen, the higgsono
couplings are rather insensitive to $\tan \bt $ (see Eqs.
(\ref{q-sq-Uhiggsino1})-(\ref{q-sq-Dhiggsino})) and therefore no
such constraints would apply. The detailed investigation of a relic
density for
 neutralino (or other candidate) cold dark matter within
proposed scenario is beyond the scope of this work.

\section{Inclusion of Light Families}

Now let us discuss the possible mass generation for the light
families. This must be done in such a way as  not to cause any
phenomenological inconsistency. Here we present one possible
 way which leads to realistic phenomenology.
We assume that all three families of the down type quarks and charged leptons
 obtain masses radiatively (similar to the
bottom quark and tau lepton) by introducing additional vector like
states near TeV scale. Thus, we introduce three vector like pairs
$(D^c+\bar D^c)_i$, $(L+\bar L)_i$ with $i=1,2,3$.
Without any selection rule, these new states may be a new source for flavor
 violation. To avoid this some care should
be exercised. Thus we postulate a flavor symmetry $SO(3)$ which will
guarantee the flavor conservation. Three families of quarks and
leptons and additional vector like pair are triplets ${\bf 3}$ of
$SO(3)$ while the higgs superfields are the singlets. Thus, all
bi-linear (mass or ${\rm mass}^2$) couplings are universal and
diagonal. For desirable fermion masses and mixings, we introduce the
flavon superfields $\xi $ and $\chi $ in the fundamental ${\bf
3}$-plet and symmetric  ${\bf 5}$-plet representations of $SO(3)$
respectively. The superpotential couplings relevant for the up type
quark sector are
\beq \fr{1}{M}(\al_MM{\bf 1}+\al_u \xi +\bt_u\chi )_{ij}q_iu^c_j\hat{h}_u~,
\la{Wup3}
\eeq
where $M$ is some cut off scale and $\al_M, \al_u, \bt_u$ some dimensionless couplings.
Clearly,  for obtaining the observed values of the masses $m_{u,c,t}$, proper selection of the
VEVs $\lan \xi \ran $, $\lan \chi \ran $ and the couplings $\al_{M, u}, \bt_u$ is needed.
For this purpose, details of $SO(3)$ flavor symmetry breaking need to be addressed (in a spirit
of Ref. \cite{Stech:2008wd}), which is beyond the scope of this work.
 Now let us turn to the down quark sector. The relevant superpotential
couplings are \beq \fr{\lam_D}{M'}(M'{\bf 1}+\al_d \xi +\bt_d\chi
)_{ij}q_iD^c_j\hat{h}_d+M_DD^c_i\bar D^c_i~, \la{Wdown3} \eeq while
the relevant soft SUSY breaking terms are \beq
m_{\tl{q}}^2|\tl{q}_i|^2+m_{\tl{d}^c}^2|\tl{d}^c_i|^2+m_{dD}^2(\tl{d}^{*c}_i\tl{D}^c_i+\tl{d}^c_i\tl{D}^{*c}_i)+
m_{\tl{D}^c}^2|\tl{D}^c_i|^2+m_{\tl{\bar D}^c}^2|\tl{\bar
D}^c_i|^2~. \la{soft3} \eeq By the loops similar to one given in
Fig. \ref{fig:1}a the operator $\propto \fr{\lam_D}{M'}(M'{\bf
1}+\al_d \xi +\bt_d\chi )_{ij}q_id^c_jh_u^{\dag }$ will be generated
which will give masses to all light down type quarks. Note the
crucial point, that all mass couplings in (\ref{Wdown3}) and
(\ref{soft3}) are degenerate. Therefore, after diagonalization of
the quark mass matrices, all these bilinear couplings will remain
diagonal and will not cause any additional flavor violation. This is
indeed the merit of the $SO(3)$ flavor symmetry in our
construction\footnote{In various constructions,  the same flavor
symmetry has been proven to be very useful for solving SUSY flavor
problem \cite{Pouliot:1993zm} and also building predictive fermion
mass pattern \cite{Stech:2008wd}.}.

The charged lepton sector can be constructed in a same manner. The relevant
 superpotential and soft SUSY breaking operators are
respectively \beq \fr{\lam_L}{M''}(M''{\bf 1}+\al_l \xi +\bt_l\chi
)_{ij}L_ie^c_j\hat{h}_d+M_LL_i\bar L_i~, \la{Wlepton3} \eeq \beq
m_{\tl{l}}^2|\tl{l}_i|^2+m_{\tl{e}^c}^2|\tl{e}^c_i|^2+m_{lL}^2(\tl{l}^{*}_i\tl{L}_i+\tl{l}_i\tl{L}^{*}_i)+
m_{\tl{L}}^2|\tl{L}_i|^2+m_{\tl{\bar L}}^2|\tl{\bar L}_i|^2~.
\la{soft-lept3}
\eeq
 One can check that the operator  $\propto
\fr{\lam_L}{M''}(M''{\bf 1}+\al_l \xi +\bt_l\chi
)_{ij}l_ie^c_jh_u^{\dag }$ will be generated by the loops similar to
one given in Fig. \ref{fig:1}b. The flavor is still conserved due to
mass operator degeneracies in (\ref{Wlepton3}) and
(\ref{soft-lept3}) insured by $SO(3)$ flavor symmetry.

Finally, let us point out that the $R$-symmetry applied and
discussed in sect. \ref{sect:FMSSM} works out also for three
families and is compatible with $SO(3)$ flavor symmetry.

\section{Embedding in $SU(5)$ GUT}

Grand unified theories have many virtues \cite{Pati:1974yy}. It is
highly motivated to extend phenomenologically interesting scenario
to the GUT. As it turns out, it is straightforward to embed the
scenario discussed in this paper in the grand unification. Let us
demonstrate this as an example for SUSY $SU(5)$. As we know, the
matter sector of minimal $SU(5)$ consists of one $(10+\bar 5)$ per
generation: $10_i=(q,u^c,e^c)_i$ and $\bar 5_i=(d^c, l)_i$
($i=1,2,3$). Since in the previous section we have seen that
inclusion of all three families of quarks and leptons is possible,
here we make presentation only with a single family. The up and down
Higgs supermultiplets $h_u$ and $h_d$ are embedded in $H(5)$ and
$\bar H(\bar 5)$ respectively. We extend the matter sector with the
vector like pair $\bar F(\bar 5)+F(5)$ with a composition: $\bar
F=(D^c, L)$, $F=(\bar D^c, \bar L)$. The up type quark mass is
generated through the superpotential coupling \beq
10\hs{-0.5mm}\cdot \hs{-0.5mm}10\hs{0.5mm} H~. \la{10-10-H}
 \eeq
This operator includes the interaction of Eq. (\ref{Yt}). In order
to realize the mechanism proposed in Section \ref{sect:FMSSM},
 one should not couple $\bar H$ directly to
the light matter (all this still can be suitably achieved by $R$-symmetry).
Therefore, down quark and charged lepton masses
should be generated radiatively. Thus we introduce the following
$SU(5)$ invariant superpotential couplings
\beq
\lam_F ~F~10\bar H+M_F\bar FF~,
\la{F-matter}
\eeq
and the following soft SUSY breaking terms
\beq
m_{10}^2|\tl{10}|^2+m_{\bar 5}|\tl{\bar 5}|^2+m_{5F}^2\l
\tl{\bar 5}\tl{\bar F}^{*}+\tl{\bar 5}^*\tl{\bar F}\r. \la{soft-SU5}
\eeq We assume that the mass $M_F$ is close to TeV scale and also
the $\mu $ term has the same magnitude. The latter  will be the case
 by achieving the doublet-triplet splitting (for instance through
the superpotential couplings $\bar H(M_H+24_H)H$ by fine tuning).
With these, we can see that all terms of (\ref{Yt})-(\ref{sbt}) can
be reproduced and the radiative generation of $m_b$ and $m_{\tau }$
can take place. For illustrative purpose we have listed the
couplings in Eqs. (\ref{10-10-H})-(\ref{soft-SU5}) in terms of
$SU(5)$ states. However, at the weak scale, the couplings and the
masses of different $SU(3)_c\tm SU(2)_L\tm U(1)_Y$ representations
will differ. This can give a desirable mismatch not only between
$m_b$ and $m_{\tau}$, but also the experimentally observed values of
$m_{\mu }/m_s$ and $m_e/m_d$ if these masses are also generated
radiatively. Together with this it would be interesting to study in
more details the properties and phenomenology of such grand unified
model. However, this is beyond the scope of this paper.

\section{Conclusions}

In this work, we proposed a new extension of MSSM  -the FMSSM - with
vector like matter at TeV scale, allowing radiative mass generation
for down type quarks and charged leptons, thus providing a plausible
explanation of why the bottom quark and the tau lepton are so much
lighter than the top quark. The $R$-symmetry which we have used
provides natural realization of the FMSSM. In particular all
couplings are controlled by this symmetry with the $\mu $ and
$B_{\mu }$ terms naturally near the TeV scale. At the same time, the
matter parity is automatic and therefore no unwanted baryon and
lepton number violating couplings are allowed. Also, the LSP is
stable, providing the desired dark matter candidate.

This scenario opens up possibility for having fermiophobic and
gaugephobic heavy Higgs sector consisting of the charged ($H^{\pm
}$) and neutral heavy Higgses ($H$ and $A$). We have discussed the
phenomenological implications of the model, which are distinct from
the usual MSSM, as well as the possible cosmological implications
for the cold dark matter.
Many of these predictions can be tested by
 upcoming experiments at the LHC.

 \vs{0.2cm}
\hs{-0.3cm}\textbf{Acknowledgement}
\vs{0.2cm}

\hs{-0.6cm}We are very grateful to D. Zeppenfeld for many helpful discussions.
SN would like to thank the warm hospitality and support of KITP,
Santa Barbara, and the organizers of the Workshop "Physics of the
Large Hadron Collider" during his participation there when this work
was completed. This work was supported in part by the US Department
of Energy, Grant Numbers DE-FG02-04ER41306 and DE-FG02-ER46140. This
research was also supported in part by the National Science
Foundation under Grant Number PHY05-51164.

\bibliographystyle{unsrt}

\end{document}